\documentstyle{article}

\begin{document}

\title{\flushright{\small KL-TH / 03-03} \bigskip \bigskip \bigskip \bigskip
\bigskip \bigskip \\
\center{ A Map between $\left( q,h \right) $-deformed Gauge Theories
                               and ordinary Gauge Theories}\thanks{%
Supported by DAAD}}
\author{L. Mesref\thanks{%
Email: lmesref@physik.uni-kl.de}}
\date{Department of Physics, Theoretical Physics\\
University of Kaiserslautern, Postfach 3049\\
67653 Kaiserslautern, Germany}
\maketitle

\begin{abstract}
We introduce a new map between a $\left( q,h\right) $-deformed gauge theory and an
ordinary gauge theory \ in a full analogy with Seiberg-Witten map.  \newpage 
\end{abstract}

\section{Introduction}

In the past few years a huge amount of literature has been devoted to the
study of quantum planes \cite{manin}. They provide a mathematical \ model
for the system under consideration. All the properties of interest can be
expressed in terms of \ noncommutative functions defined on these quantum
spaces. Quantum groups \cite{drinfeld} are genetrated by such functions. It
was shown that the only quantum groups which preserve nondegenerate bilinear
forms are $GL_{qp}\left( 2\right) $ and $GL_{hh^{\prime }}\left( 2\right) $,
\cite{demidov}-\cite{aghamohammadi}. They act on the $q$-plane (Manin
plane), with relation $\hat{x}\hat{y}=q\hat{y}\hat{x}$ and on the $h$-plane (Jordanian plane), with
relation $\hat{x}\hat{y}-\hat{y}\hat{x}=h\hat{y}^{2}$, respectively.

In a recent paper \cite{mesref1} we have constructed a map relating $q$%
-deformed gauge fields defined on the Manin plane and ordinary gauge fields
in a full analogy with Seiberg-Witten map \cite{seiberg}. This work has been
extended to $GL_{q}\left( N\right) $-covariant quatum hyperplane \cite
{mesref2}. In the present letter we define a map which relates $\left(
q,h\right) $-deformed gauge fields defined on the hybrid quantum space (
given by $\hat{x}\hat{y}-q\hat{y}\hat{x}=h\hat{y}^{2}$) and the ordinary gauge fields. The product of
functions on such a space is obtained by applying a singular transformation
\cite{aghamohammadi} on the Gerstenhaber star product \cite{gerstenhaber}.

\section{$\left( q,h \right) $-deformed gauge theory versus ordinary gauge theory}

To begin we consider the undeformed action

\bigskip

\begin{equation}
S=-\frac{1}{4}\int d^{4}x\,\,F_{\mu \nu }F^{\mu \nu },  
\end{equation}

\bigskip

where

\bigskip

\begin{equation}
F_{\mu \nu }=\partial _{\mu }A_{\nu }-\partial _{\nu }A_{\mu }. 
\end{equation}

\bigskip

$S$ is invariant with respect to infinitesimal gauge transformations:

\bigskip

\begin{equation}
\delta _{\lambda }A_{\mu }=\partial _{\mu }\lambda .  
\end{equation}

\bigskip

To study the $\left( q,h\right) $-deformed analogue of this gauge theory let
us first recall that the Manin plane, with the relation $\hat{X}\hat{Y}=q\hat{Y}\hat{X}$ and the
hybrid plane with the relation $\hat{x}\hat{y}-q\hat{y}\hat{x}=h\hat{y}^{2}$ are related by a
transformation \cite{aghamohammadi}

\bigskip

\begin{eqnarray}
\left( 
\begin{array}{c}
\hat{X} \\ 
\hat{Y}
\end{array}
\right) &=&\left(
\begin{array}{cc}
1 & \alpha \\
0 & 1
\end{array}
\right) \left( 
\begin{array}{c}
\hat{x} \\ 
\hat{y}
\end{array}
\right) ,  \nonumber \\
\left( 
\begin{array}{c}
\partial _{\hat{X}} \\
\partial _{\hat{Y}}
\end{array}
\right) &=&\left(
\begin{array}{cc}
1 & 0 \\
-\alpha & 1
\end{array}
\right) \left(
\begin{array}{c}
\partial _{\hat{x}} \\
\partial _{\hat{y}}
\end{array}
\right) ,  
\end{eqnarray}

\bigskip

where $\alpha =\frac{h}{q-1}$.

The product of functions is defined via the Gerstenhaber star product \cite
{gerstenhaber}: Let $\mathcal{A}$ be an associative algebra and let $%
D_{i},E^{i}:\mathcal{A}\rightarrow \mathcal{A}$ be a pairwise derivations.
Then the associative star product of $a$ and $b$ is given by 

\bigskip 

\begin{equation}
a\star b=\mu \circ e^{\zeta \sum_{i}D_{i}\otimes E^{i}}a\otimes b,  
\end{equation}

\bigskip 

where $\zeta $ is a parameter and $\mu $ the undeformed product given by

\bigskip

\begin{equation}
\mu \left( f\otimes g\right) =fg.  
\end{equation}

\bigskip 

On the Manin plane, we write this star product as:

\bigskip

\begin{equation}
f\star g=\mu \circ e^{\frac{i\eta }{2}\left( X\frac{\partial }{\partial X}%
\otimes Y\frac{\partial }{\partial Y}-Y\frac{\partial }{\partial Y}\otimes X%
\frac{\partial }{\partial X}\right) }f\otimes g. 
\end{equation}

\bigskip 

A straightforward computation gives then the following commutation relations

\bigskip

\begin{equation}
X\star Y=e^{\frac{i\eta }{2}}XY,\qquad \qquad Y\star X=e^{\frac{-i\eta }{2}%
}YX.  
\end{equation}

\bigskip

Whence

\bigskip

\begin{equation}
X\star Y=e^{i\eta }Y\star X,\qquad q=e^{i\eta }. 
\end{equation}

\bigskip

Thus we recover the commutation relations for the Manin plane.

We can also write the product of functions as

\bigskip

\begin{equation}
f\star g=fe^{\frac{i}{2}\overleftarrow{\partial }_{k}\Theta ^{kl}\left(
X,Y\right) \overrightarrow{\partial }_{l}}g 
\end{equation}

\bigskip

where $\Theta ^{kl}\left( X,Y\right) =\eta XY\epsilon ^{kl}$ with $\epsilon
^{12}=-\epsilon ^{21}=1$.

Using (4) we can define the product of functions on the hybrid space as

\bigskip

\begin{equation}
f\star g=\mu \circ e^{\frac{i\eta }{2}\left[ \left( x\frac{\partial }{%
\partial x}+\alpha y\frac{\partial }{\partial x}\right) \otimes \left( y%
\frac{\partial }{\partial y}-\alpha y\frac{\partial }{\partial x}\right)
-\left( y\frac{\partial }{\partial y}-\alpha y\frac{\partial }{\partial x}%
\right) \otimes \left( x\frac{\partial }{\partial x}+\alpha y\frac{\partial 
}{\partial x}\right) \right] }\left( f\otimes g\right) .  
\end{equation}

\bigskip 

A direct computation gives

\bigskip

\begin{equation}
x\star y=e^{\frac{i\eta }{2}}xy+\left( e^{\frac{i\eta }{2}}-1\right) \alpha
y^{2},\quad y\star x=e^{\frac{-i\eta }{2}}yx+\left( e^{\frac{-i\eta }{2}%
}-1\right) \alpha y^{2},  
\end{equation}

\bigskip

whence

\bigskip

\begin{equation}
x\star y=e^{i\eta }y\star x+\left( e^{i\eta }-1\right) \alpha y^{2}.
\end{equation}

\bigskip

Thus we recover the commutation relations for the hybrid plane $x\star
y=qy\star x+hy^{2}$ with $q=e^{i\eta }$.

Expanding to first nontrivial order in $\eta $ and $h$ $\left( \eta
,h<<1\right) $ we find

\bigskip

\begin{eqnarray}
f\star g &=&fg+\frac{i\eta }{2}\left( x\frac{\partial f}{\partial x}+\alpha y%
\frac{\partial f}{\partial x}\right) \left( y\frac{\partial g}{\partial y}%
-\alpha y\frac{\partial g}{\partial x}\right)   \nonumber \\
&&-\frac{i\eta }{2}\left( y\frac{\partial f}{\partial y}-\alpha y\frac{%
\partial f}{\partial x}\right) \left( x\frac{\partial g}{\partial x}+\alpha y%
\frac{\partial g}{\partial x}\right)   \nonumber \\
&=&fg+\frac{i}{2}\left( \eta xy-ihy^{2}\right) \left( \frac{\partial f}{%
\partial x}\frac{\partial g}{\partial y}-\frac{\partial f}{\partial y}\frac{%
\partial g}{\partial x}\right) .  
\end{eqnarray}

We can also write (11) as

\bigskip

\begin{equation}
f\star g=f\,\,e^{\frac{i}{2}\overleftarrow{\partial }_{k}\theta ^{kl}\left(
x,y\right) \overrightarrow{\partial }_{l}}\,g.  
\end{equation}

\bigskip

Here the antisymmetric matrix $\ \theta ^{kl}\left( x,y\right) =\eta \left(
x+\alpha y\right) y\epsilon ^{kl}=\left( \eta xy-ihy^{2}\right) \epsilon
^{kl}$ with $\epsilon ^{12}=-\epsilon ^{21}=1$.

The $\left( q,h\right) $-deformed infinitesimal gauge transformations are
defined by

\bigskip

\begin{eqnarray}
\widehat{\delta }_{\widehat{\lambda }}\widehat{A}_{\mu } &=&\partial _{\mu }%
\widehat{\lambda }+i\left[ \widehat{\alpha },\widehat{A}_{\mu }\right]
_{\star }=\partial _{\mu }\widehat{\lambda }+i\widehat{\lambda }\star 
\widehat{A}_{\mu }-i\widehat{A}_{\mu }\star \widehat{\lambda },  \nonumber \\
\widehat{\delta }_{\widehat{\lambda }}\widehat{F}_{\mu \nu } &=&i\widehat{%
\lambda }\star \widehat{F}_{\mu \nu }-i\widehat{F}_{\mu \nu }\star \widehat{%
\lambda }.  
\end{eqnarray}

\bigskip

To first order in $\theta $, the above formulas for the gauge
transformations read

\bigskip

\begin{eqnarray}
\widehat{\delta }_{\widehat{\lambda }}\widehat{A}_{\mu } &=&\partial _{\mu }%
\widehat{\lambda }-\frac{1}{2}\theta ^{\rho \sigma }\left( x,y\right) \left(
\partial _{\rho }\lambda \partial _{\sigma }A_{\mu }-\partial _{\rho }A_{\mu
}\partial _{\sigma }\lambda \right) ,  \nonumber \\
\widehat{\delta }_{\widehat{\lambda }}\widehat{F}_{\mu \nu } &=&-\frac{1}{2}%
\theta ^{\rho \sigma }\left( x,y\right) \left( \partial _{\rho }\lambda
\partial _{\sigma }F_{\mu \nu }-\partial _{\rho }F_{\mu \nu }\partial
_{\sigma }\lambda \right) .  
\end{eqnarray}

\bigskip

To ensure that an ordinary gauge transformation of $A$ by $\lambda $ is
equivalent to $\left( q,h\right) $-deformed gauge transformation of $%
\widehat{A}$ by $\widehat{\lambda }$ we consider the following relation \cite
{seiberg}

\bigskip

\begin{equation}
\widehat{A}\left( A\right) +\widehat{\delta }_{\widehat{\lambda }}\widehat{A}%
\left( A\right) =\widehat{A}\left( A+\delta _{\lambda }A\right) .  
\end{equation}

\bigskip

We first work the first order in $\theta $

\bigskip

\begin{eqnarray}
\widehat{A} &=&A+A^{\prime }\left( A\right)   \nonumber \\
\widehat{\lambda }\left( \lambda ,A\right)  &=&\lambda +\lambda ^{\prime
}\left( \lambda ,A\right) .  
\end{eqnarray}

Expanding (18) in powers of $\theta $ we find

\bigskip

\begin{equation}
A_{\mu }^{\prime }\left( A+\delta _{\lambda }A\right) -A_{\mu }^{\prime
}\left( A\right) -\partial _{\mu }\lambda ^{\prime }=\theta ^{kl}\left(
x,y\right) \partial _{k}A_{\mu }\partial _{l}\lambda .
\end{equation}

\bigskip

The solutions are given by

\bigskip

\begin{eqnarray}
\widehat{A}_{\mu } &=&A_{\mu }-\frac{1}{2}\theta ^{\rho \sigma }\left(
x,y\right) \left( A_{\rho }F_{\sigma \mu }+A_{\rho }\partial _{\sigma
}A_{\mu }\right) ,  \nonumber \\
\widehat{\lambda } &=&\lambda +\frac{1}{2}\theta ^{\rho \sigma }\left(
x,y\right) A_{\sigma }\partial _{\rho }\lambda . 
\end{eqnarray}

\bigskip

The $q$-deformed curvature $\widehat{F}_{\mu \nu }$ is given by

\bigskip

\begin{eqnarray}
\widehat{F}_{\mu \nu } &=&\partial _{\mu }\widehat{A}_{\nu }-\partial _{\nu }%
\widehat{A}_{\mu }-i\left[ \widehat{A}_{\mu },\widehat{A}_{\nu }\right]
_{\star }  \nonumber \\
&=&\partial _{\mu }\widehat{A}_{\nu }-\partial _{\nu }\widehat{A}_{\mu }-i%
\widehat{A}_{\mu }\star \widehat{A}_{\nu }+i\widehat{A}_{\nu }\star \widehat{%
A}_{\mu }.  
\end{eqnarray}

\bigskip

Using (21) we find

\bigskip

\begin{eqnarray}
\widehat{F}_{\mu \nu } &=&F_{\mu \nu }+\theta ^{\rho \sigma }\left( x\right)
\left( F_{\mu \rho }F_{\nu \sigma }-A_{\rho }\partial _{\sigma }F_{\mu \nu
}\right)  \nonumber \\
&&-\frac{1}{2}\partial _{\mu }\theta ^{\rho \sigma }\left( x\right) \left(
A_{\rho }F_{\sigma \nu }+A_{\rho }\partial _{\sigma }A_{\nu }\right)
\nonumber \\
&&+\frac{1}{2}\partial _{\nu }\theta ^{\rho \sigma }\left( x\right) \left(
A_{\rho }F_{\sigma \mu }+A_{\rho }\partial _{\sigma }A_{\mu }\right) , 
\end{eqnarray}

\bigskip

which we can write as

\bigskip

\begin{equation}
\widehat{F}_{\mu \nu }=F_{\mu \nu }+f_{\mu \nu }+o\left( \theta ^{2}\right) ,
\end{equation}

\bigskip

where $f_{\mu \nu }$ is the quantum correction linear in $\theta $. The
quantum analogue of the action (1) is given by

\bigskip

\begin{equation}
\widehat{S}=-\frac{1}{4}\int d^{4}x\,\,\widehat{F}_{\mu \nu }\star \widehat{F%
}^{\mu \nu }.  
\end{equation}

\bigskip

For functions $f$, $g$ that vanish rapidly enough at infinity, the measure is defined such that

\bigskip

\begin{equation}
\int f \star g = \int g \star f =\int f \,\,\, g.
\end{equation}

\bigskip

From this equation we sse that the kinetic part the action is the same as its commutative version. So the free field propagators in commutative and non commutative spaces are the same.

The deformed action (25) reads

\bigskip

\begin{equation}
\widehat{S}=S+S_{\theta },  
\end{equation}

\bigskip

where $S$ is the undeformed action (1) and $S_{\theta }$ is the correction linear in $\theta $.

\bigskip

Now, we make the following remarks. First for $h=0\,\,\left( \theta
^{kl}\left( x,y\right) =\eta xy\epsilon ^{kl}\right) $, the action (25) is
just the first order $q$-deformed action.

Second, for $\eta =0\,\,\left( \theta ^{kl}\left( x,y\right)
=-ihy^{2}\epsilon ^{kl}\right) $, the action (25) is the Jordanian $h$%
-deformed action.

For $h=0$ and $q=1\,\left( \eta =0\right) $ this action coincides with the
classical action (1) as it should be.

\bigskip

To conclude, we have obtained a more genereral Seiberg-Witten map in $\left(
q,h\right) $-deformed quantum plane which reduces to the known ones at some
limits. The method developed in this letter can be applied to various
quantum deformed models \cite{mesref3}-\cite{mesref6}.

\bigskip

\end{document}